\documentstyle[epsf]{lamuphys}
\makeatletter
\let\chapter\hid@chapter
\makeatother
\begin{document}
\pagenumbering{arabic}
\title{A Complete 2dF Survey of Fornax}

\author{M.J.\,Drinkwater\inst{1}, E.M.\,Sadler\inst{2}, 
J.I.\,Davies\inst{3},
R.J.\,Dickens\inst{4},
M.D.\,Gregg\inst{5},\\
Q.A.\,Parker\inst{6},
S.\,Phillipps\inst{4}, 
R.M.\,Smith\inst{3}}

\institute{University of New South Wales, Physics, Sydney 2052, Australia
\and
University of Sydney, Physics, NSW 2006, Australia
\and
University of Wales, Cardiff, Physics, PO Box 913, Cardiff CF2 3YB, UK
\and
University of Bristol, H.H.\,Wills Physics Lab., Tyndall Av., Bristol BS8 1TL, UK
\and
IGPP, Lawrence Livermore National Laboratory, Livermore, CA 94550, USA
\and
Anglo-Australian Observatory, Coonabarabran, NSW 2357, Australia
}
\authorrunning{M.\,Drinkwater et al.}
\maketitle

\begin{abstract}
We are using the 2dF spectrograph on the Anglo-Australian Telescope
to obtain spectra for a complete sample of all 14000 objects
with $16.5<B<19.7$ in a 12 square degree area centred on the Fornax
cluster. The aims of this project include the study of dwarf galaxies
in the cluster (both known low surface brightness objects and putative
normal surface brightness dwarfs) and a comparison sample of
background field galaxies. We will also measure quasars, any
previously unrecognised compact galaxies and a large sample of
Galactic stars.  Here we present initial results from the first 680
objects observed, including the discovery of a number of dwarf
galaxies in the cluster more compact than any previously known.
\end{abstract}

\section{The Complete Sample}

Our primary goal is to obtain a complete sample of galaxies over a
large range of magnitude and surface brightness to study the
luminosity function and dynamics of both the Fornax cluster and
background galaxies. Previous cluster samples were compiled from 2-D
images without spectra so it was hard to tell if small galaxies were
cluster dwarfs or background giants.  We can solve this problem with
2dF which allows us to make a complete spectroscopic survey in the
direction of the Fornax cluster. A further limitation of most existing
galaxy surveys is that they only considered resolved images, so were
biased against compact galaxies.  Our survey will measure all images
and thus avoids this bias.
 
One important advantage of our survey is that by including all
morphological types it will provide a unique test of the presumed
continuity of QSOs and Seyfert-1s as well as measuring the Seyfert
luminosity function.  Our 2dF spectra have a resolution of 0.85\,nm
(400 km/s) and can therefore resolve the broad lines of active
galaxies. The only other unbiased Seyfert samples have been limited to
very bright magnitudes, (e.g. the Hamburg QSO survey to $B<17$, Kohler
et al.\ 1997). Our sample of 12 square degrees to $B<19.7$ will
contain 200 QSOs and 50 Seyfert 1s, all with spectral classifications.
\raisebox{-3cm}[0cm][0cm]{\makebox[0cm][c]
{\parbox{12cm}{\em Proceedings of the ESO/ATNF
Workshop ``Looking Deep in the Southern Sky'',
Sydney, Australia, 10--12 December 1997, eds. R. Morganti and W. Couch
}}}

In this paper we describe the results of our first 2dF observations of
a sample of 300 galaxies (1 and 2 hour exposures) and 380 unresolved
sources (30 minute exposure).  The 380 stellar sources were chosen with
a bias to very blue and very red stars.

\subsection{Galaxy Sample Results}

Our galaxy observations have confirmed many members of the cluster and
we have also discovered 7 new dwarf cluster galaxies; these are among
the most compact dwarf galaxies known (see Drinkwater \& Gregg,
1998). Three of the new cluster members show strong emission lines and
are very small blue compact dwarf galaxies and one of these may be the
first true dwarf spiral discovered.  We are also correlating our
optical data with an 843 MHz radio continuum survey of the field with
the University of Sydney MOST telescope. Two radio sources we have
identified with quasars are shown in Fig.~1.
 
\begin{figure}
\centering
\epsfxsize=12.8cm\epsfbox[42 160 592 370]{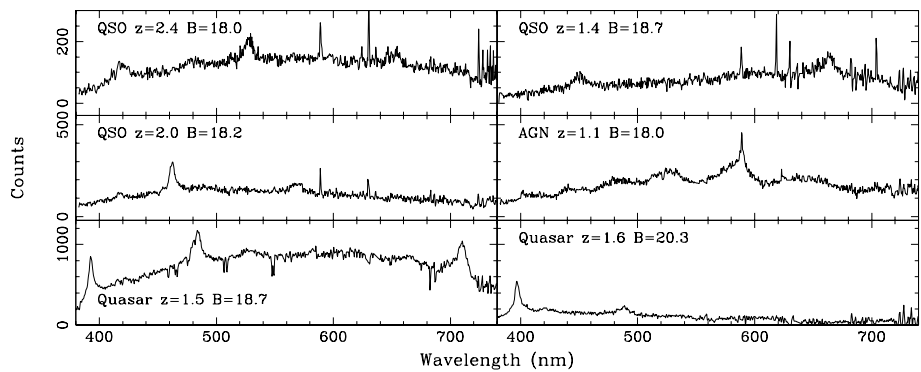}
\caption{2dF spectra of blue stellar objects (upper 4 panels, 30 min
exposure) and radio-loud sources (lower 2 panels, 2 h exposure).}
\end{figure}
 
\subsection{Stellar Sample Results}


Many of the bluest stellar images are QSOs; we detected 13 (see
Fig.~1). Allowing for the fraction of all stars observed and our
magnitude limit, this number is quite consistent with the expected QSO
number counts (for $z<2.2$ and $B<19.75$) of 55 per 2dF (Boyle et al.\
1990).  We found one unusual AGN spectrum in the stellar sample, shown
in Fig.~1. The broad bands are real, although the peak at 590\,nm is
next to a poorly removed night sky emission line. This source has
previously been identified as an X-ray source at a red shift of
$z=1.1$ from the Einstein Medium Sensitivity Survey (Stoke et al.\
1991).

%

\end{document}